\documentclass[twocolumn,epjc3]{svjour3}
\pdfoutput=1

\usepackage[latin1]{inputenc}
\usepackage{graphicx}
\usepackage{xspace}

\usepackage{multirow}
\usepackage{textcomp}

\renewcommand{\b}{{\mathrm b}}
\renewcommand{\c}{{\mathrm c}}

\newcommand{\e}{{\mathrm e}}

\newcommand{\g}{{\mathrm g}}

\newcommand{\p}{{\mathrm p}}
\newcommand{\q}{{\mathrm q}}

\renewcommand{\t}{{\mathrm t}}
\renewcommand{\u}{{\mathrm u}}

\newcommand{\B}{{\mathrm B}}

\renewcommand{\H}{{\mathrm H}}
\newcommand{\J}{{\mathrm J}}
\newcommand{\K}{{\mathrm K}}

\newcommand{\W}{{\mathrm W}}
\newcommand{\Z}{{\mathrm Z}}
\newcommand{\bbar}{\overline{\mathrm b}}
\newcommand{\cbar}{\overline{\mathrm c}}

\newcommand{\qbar}{\overline{\mathrm q}}

\newcommand{\sbar}{\overline{\mathrm s}}
\newcommand{\tbar}{\overline{\mathrm t}}

{\end{list}}
\newcounter{enumct}

\newlength{\abstwidth}
\newcommand{\secRef}[1]{section~\ref{#1}\xspace}

\newcommand{\tabRef}[1]{tab.~\ref{#1}\xspace}

\newcommand{\figRef}[1]{fig.~\ref{#1}\xspace}

\title{\LARGE\bf Colour Reconnection at Future e$^{\mathbf{+}}$e$^{\mathbf{-}}$ 
Colliders}

\author{Jesper R. Christiansen\thanksref{e1,addr1}
        \and
        Torbj\"orn Sj\"ostrand\thanksref{e2,addr1}}

\thankstext{e1}{e-mail: jesper.christiansen@thep.lu.se}
\thankstext{e2}{e-mail: torbjorn@thep.lu.se}
\institute{Theoretical High Energy Physics, 
   Department of Astronomy and Theoretical Physics, 
   Lund University, 
   S\"olvegatan 14A, SE-223 62 Lund, Sweden \label{addr1}}
 \date{June 2015}
 
\begin{document}
\sloppy

\maketitle
%
%

\vspace{\fill}

\begin{abstract}
The effects of colour reconnection (CR) at $\e^+\e^-$ colliders are 
revisited, with focus on recently developed CR models. The new models 
are compared with the LEP2 measurements for 
$\e^+\e^- \to \W^+\W^- \to \q_1 \qbar_2 \q_3 \qbar_4$ and found to 
lie within their limits. Prospects for constraints from new 
high-luminosity $\e^+\e^-$ colliders are discussed. The novel arena of
CR in Higgs decays is introduced, and illustrated by shifts in angular
correlations that would be used to set limits on a potential $CP$-odd 
admixture of the 125 GeV Higgs state.
\end{abstract}





\section{Introduction}

Multiparticle production in high-energy collisions often involves many
contributing intermediate sub-sources. The cleanest such example is 
$\e^+\e^- \to \W^+\W^- \to \q_1 \qbar_2 \q_3 \qbar_4$, or its equivalent 
with a $(\gamma^*/\Z^0)(\gamma^*/\Z^0)$ intermediate state. A more tricky 
one is multiparton interactions (MPIs) in hadronic collisions, wherein
a variable set of (semi)perturbative partonic collisions together with 
the beam remnants are at the origin of the subsequent hadronization.

In neither case can a first-principles QCD calculation be carried out
to describe the particle production process. Instead string or cluster
models are used \cite{Buckley:2011ms}. Both are based on an $N_C \to \infty$ 
limit \cite{'tHooft:1973jz}, wherein each colour-anticolour
pair is unique. Thus, in the string model, each quark is at the end 
of a string, whereas a gluon is attached to two string pieces and thus
forms a kink on a longer string usually stretched between an endpoint 
quark and ditto antiquark \cite{Andersson:1983ia}. In simple systems like    
$\e^+\e^- \to \gamma^*/\Z^0 \to \q\qbar\g$ such principles give unique 
topologies, but for more complicated situations ambiguities arise.  
When these can be associated with the presence of unexpected colour 
topologies we speak of colour reconnection (CR). 
The historical example in this spirit is the decay
$\B^+ = \u \bbar \to \u \cbar \W^+ \to (\u \cbar) (\c \sbar) \to 
(\u \sbar)(\c\cbar) \to \K^+ \, \J/\psi \to \K^+\mu^+\mu^-$ 
\cite{Fritzsch:1979zt}, where we have used brackets in intermediate 
states to delineate separate colour singlet identities. 

Similarly, for $\e^+\e^- \to \W^+\W^-$, with $\W^+ \to \q_1 \qbar_2$ and 
$\W^- \to \q_3 \qbar_4$, to first approximation the $\q_1 \qbar_2$ and 
$\q_3 \qbar_4$ systems hadronize separately from each other. Deviations 
from such a production picture could be parametrized as an admixture of 
alternative colour-reconnected $\q_1 \qbar_4$ and $\q_3 \qbar_2$ systems. 
Such CR was highly relevant in the context of the $\W$ mass
measurement at LEP2~\cite{Sjostrand:1993hi,Sjostrand:1993rb}, where a 
potentially non-negligible uncertainty was predicted. This led to the 
development of dedicated studies aimed directly at measuring CR in 
hadronic $\W^+\W^-$ 
events~\cite{Abbiendi:2005es,Achard:2003pe,Abdallah:2006uq,Schael:2006mz}. 
The most extreme CR models could be ruled out, but not enough statistics 
was collected to definitely distinguish between the more moderate CR models 
and no CR~\cite{Schael:2013ita}. Nevertheless such moderate-model reconnection 
in about half of all events provided the best overall description.

Modelling and testing of CR in hadronic collisions is rather more 
complicated \cite{Sjostrand:1987su,Sjostrand:2004pf}. And yet the 
case for it playing an important role is compelling, e.g.\ from the 
rise of the average transverse momentum with increasing charged 
multiplicity. Thus, given the predominance of hadronic colliders in 
recent years, first with the Tevatron and now with the LHC, recent CR 
studies have rather aimed to address the more complicated issues arising 
there, and has led to the introduction of several new 
models~\cite{Argyropoulos:2014zoa,Christiansen:2015yqa}.  
These rely only on the distribution of final state partons just prior to 
the hadronization, making them directly applicable also to $\e^+\e^-$ 
colliders. And even if the CR effects are expected to be significantly 
smaller in  $\e^+\e^-$ than in $\p\p$, this is compensated by a cleaner 
environment allowing for higher precision. On the one hand, it is 
therefore highly relevant to go back and check whether the newly developed 
models are consistent with the LEP2 data. On the other hand, it is useful
to consider what further tests may come in the future. As an example, the 
recently suggested 100~km $\e^+\e^-$ collider~\cite{Gomez-Ceballos:2013zzn}
would produce $\mathcal{O} (10^8)$ $\W^+\W^-$ pairs, resulting in 
a statistical uncertainty on the $\W$ mass below 1~MeV, e.g.\ from
semileptonic decays $\e^+\e^- \to \W^+\W^- \to \q_1 \qbar_2 \ell \nu_{\ell}$.
With the calculated mass shifts in the original
CR paper of the order 10-20 MeV~\cite{Sjostrand:1993hi} as a reference,
such a precision should make it possible to rule out many CR models,
and also (hopefully) definitely confirm the presence of CR effects.

With the discovery of the Higgs boson~\cite{Chatrchyan:2012ufa,Aad:2012tfa},
a new arena for CR studies opens up. The Higgs state is very narrow ---
the expected width is of the order of 4~MeV --- meaning that it is very 
long-lived. Therefore hadronization of the rest of the event already 
happened and the produced hadrons already spread out by the time the 
Higgs decays. That is, the Higgs itself decays essentially in a vacuum, 
and has no interactions with the rest of the event, be that in $\e^+\e^-$ 
or $\p\p$ collisions. Among its key decay channels we find $\W^+\W^-$ and 
$\Z^0\Z^0$, however, and here history repeats itself: fully hadronic 
decays would be sensitive to CR between the two gauge-boson systems. 
The variables of interest here are not only masses but even more the 
angles between the four hadronic jets. Such angles can be modified by CR,
a phenomenon which was noted e.g.\ in the context of top mass
studies~\cite{Argyropoulos:2014zoa}. CR uncertainties thereby affect 
precision measurements of the Higgs properties, one of the primary 
purposes of future $e^+e^-$ colliders. To be specific, the SM predicts 
the Higgs to be a $CP$-even state, which is also observed to be strongly 
favoured compared with the $CP$-odd 
alternative~\cite{Aad:2015mxa,Chatrchyan:2013mxa}. Extensions of the 
SM Higgs sector, however, allows for the observed Higgs to be a mixture
of both possibilities. One place to search for deviations from the predicted
SM Higgs behaviour is precisely the angular correlations in 
hadronic $\W^+\W^-$ (or $\Z^0\Z^0$) decays \cite{Skjold:1993jd}. Hence CR 
could introduce a systematic uncertainty, and in this article we do a
first study on the size of such uncertainties in various CR scenarios.

This paper is organized as follows. The different CR models we will compare
are briefly summarized in section \ref{sec:model}. The three next sections 
contain studies on three different sets of observables, namely, the 
$\W$ mass measurement, section \ref{sec:mass}, the search for CR effects
in $\W^+\W^-$ events, section \ref{sec:angular}, and the Higgs $CP$ 
measurements, section \ref{sec:higgs}. The article ends with a few 
conclusions, section \ref{sec:conclusions}. 

\section{The CR models \label{sec:model}}

Our current understanding of QCD does not provide a unique recipe for CR.
Therefore the best we can do is contrast different plausible scenarios,
and let data be the judge what works and what does not. In this article we
will compare four different CR models, which provide a reasonable spread 
of properties and predictions. Before briefly presenting each of these 
models it is useful to outline some of the basic issues that are involved. 

One key aspect is what role is given to colour algebra. To illustrate 
this, again consider 
$\e^+\e^- \to \W^+\W^- \to \q_1 \qbar_2 \q_3 \qbar_4$. From the onset,
$\q_1 \qbar_2$ form one singlet and $\q_3 \qbar_4$ another. In addition,
there is a $1/9$ probability that $\q_1 \qbar_4$ and $\q_3 \qbar_2$ 
``accidentally'' form singlets. In some models such accidental matches 
are a prerequisite to allow a CR. In this sense, these models are not 
really about \textit{re}connections but about a choice between already 
existing singlets. The alternative is to view CR as a dynamical process, 
wherein (infinitely) soft gluons can mediate any colour exchange required 
to form new singlets. The original non-accidental singlets define an 
initial state that actively needs to be perturbed to create alternative 
colour topologies. As so often, these two pictures may be viewed as
extremes, and the ``true'' behaviour may well be in between, with a 
bit of each.

Here another aspect enters, namely the role of geometry/causality. With a 
$c\tau \approx 0.1$~fm, the $\W^{\pm}$ decays tend to be separated
on a scale an order of magnitude below the typical hadronic size,
the latter also being the size of the colour fields stretched between 
colour-connected partons. It would thereby seem that the $\W^+$ and
$\W^-$ colour fields fully overlap, at least in the threshold region
where the $\W$'s are not too strongly boosted apart. Introducing causality,
however, the colour fields take some time to grow to full size (e.g. in 
the SK-I model described later). Meanwhile
they drift apart, thereby only partly overlapping, and with an overlap that
depends on the motion of all the string pieces from each $\W$ decay.
In models where geometry is allowed to play a role there is also a natural 
decoupling of the two $\W$ decays at energies well above the threshold 
region, or if the $\W$ width could be sent to zero, and this should not 
be spoiled by the ``accidental'' singlets.

Finally there is also a selection principle: if there are many potential
reconnections in an event, which are the one(s) that actually occur?
This could be at random or involve some bias. The most common bias is 
to make use of the $\lambda$ measure, which characterizes the total 
string length \cite{Andersson:1985qr}. That is, the smaller the $\lambda$,
the better ordered are the partons along the strings. The full $\lambda$
expression is rather messy, so a commonly used approximation is 
\begin{equation}
\lambda = \sum_{ij} \, \ln \left( 1 + \frac{m_{ij}^2}{m_0^2} \right) ~,
\end{equation} 
where the $ij$ sum runs over all parton pairs connected by a string piece
and $m_0$ is of the order of a typical hadronic mass.
The average hadronic multiplicity of a string piece grows roughly 
logarithmically with its mass, so a reduction of $\lambda$ corresponds
to a reduction of the ``free energy'' available for particle production.

Among the four different CR models considered in this study, SK-I and
SK-II were developed for $\W$ mass uncertainty studies at 
LEP2~\cite{Sjostrand:1993hi}. The gluon move model, GM, was
introduced as a simple model, among a few others, to study the 
effect of CR in top decays~\cite{Argyropoulos:2014zoa}. Finally,
the QCD-based model, CS, was introduced to look for effects in 
soft QCD, especially baryon production~\cite{Christiansen:2015yqa}. 
The first two models are only applicable for the hadronic decays in 
diboson production, whereas the latter two could be used for any process. 
All of the models are available in (recent versions of)
\textsc{Pythia}~8~\cite{Sjostrand:2014zea}, the first two having been 
(re)implemented expressly for this study. That program also contains
another CR model~\cite{Sjostrand:2004pf}, used by default, that relies 
on the MPI structure of hadron collisions and therefore cannot be used 
in $\e^+\e^-$. All of the algorithms are applied after the hard 
primary process and the subsequent parton-shower evolution, but before  
the hadronization step. Typically this means that each $\W$ contains a
handful of gluons, in addition to the primary $\q\qbar$ pair, when CR is 
to be considered.

Both the SK-I and SK-II model utilize the space-time picture of strings
being stretched between the different decay products of the two bosons. 
A reconnection between two string pieces from different bosons is allowed
only when these overlap in their space--time motion. Since such an overlap 
is assumed associated with the possibility for dynamical soft-gluon 
exchange between the two overlapping colour fields, there is no 
colour-factor suppression for reconnection. The two approaches
differ in their definition of what an overlap means, taking two extreme
limits by analogy with Type I and Type II superconductors, which explains
their names. In SK-I the strings are imagined as elongated bags, and the 
probability for a reconnection is proportional to the integrated 
space--time overlap between two string pieces. (Up to saturation effects 
to ensure that probabilities stay below unity.) This model contains one 
parameter that directly controls the overall strength of the CR, which 
made it convenient for experimental LEP2 studies. For SK-II the string is 
considered to contain a thin core, a vortex line, where all the 
topological information is stored, even if the full energy still is 
spread over a larger volume. A reconnection can only occur when the 
space--time motion makes two such cores cross each other. This model 
introduces no special parameters, and therefore gives unique predictions. 
(In both models one parameter is used to describe how the strings decay 
exponentially in proper time, and in SK-I additionally the string width
is a parameter, but these parameters are almost completely fixed within 
the string model itself.) Normally only one reconnection is made, namely
the one that happens first in proper time. By default this reconnection 
may either increase or decrease the total $\lambda$ measure, but in the 
primed variants SK-I$'$ and SK-II$'$ only reconnections that reduce 
$\lambda$ are considered. The SK models were tested at LEP2, where only 
the most extreme versions of SK-I were ruled out. For the SK-I model best 
agreement with data was obtained with parameter such that approximately
50\% of all events contain a reconnection, as already mentioned.

\begin{figure*}[tp]
  \centering
  \includegraphics[scale=0.38]{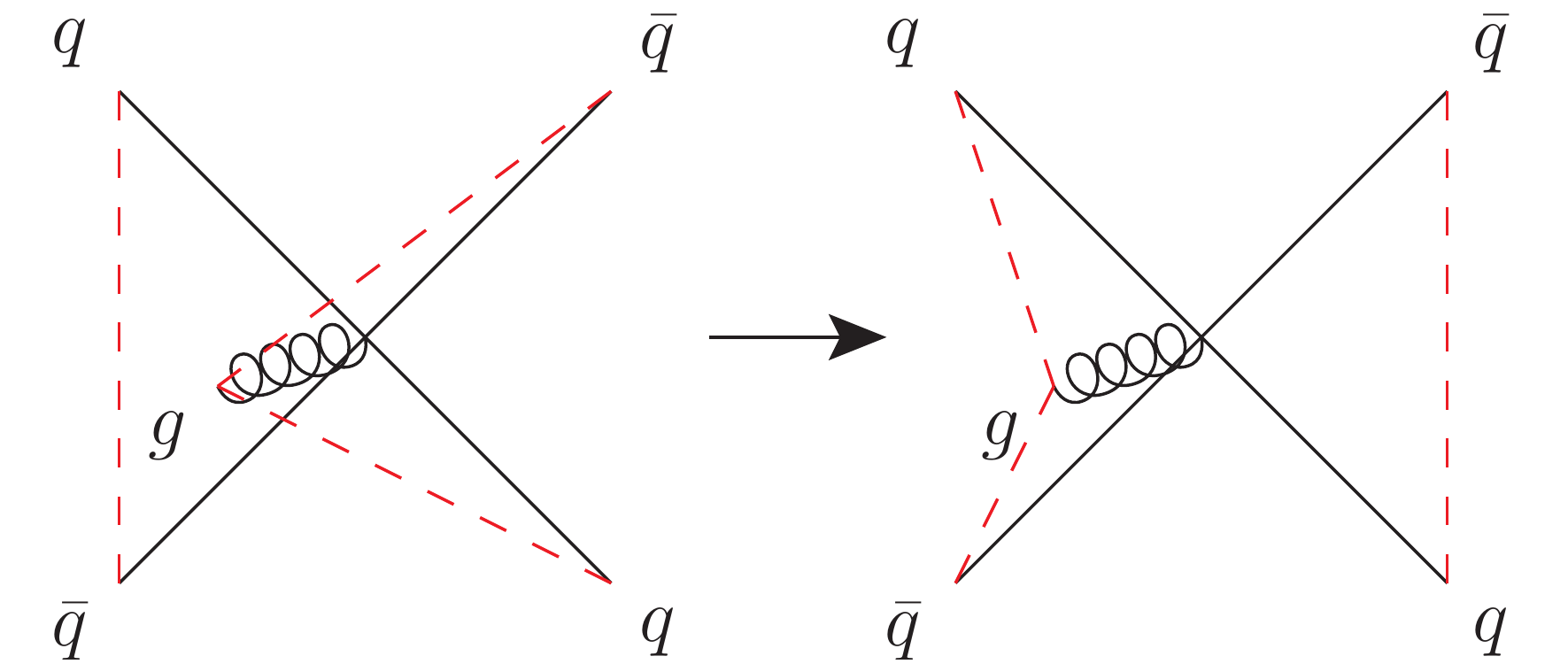}
  \hspace*{0.5cm}
  \includegraphics[scale=0.38]{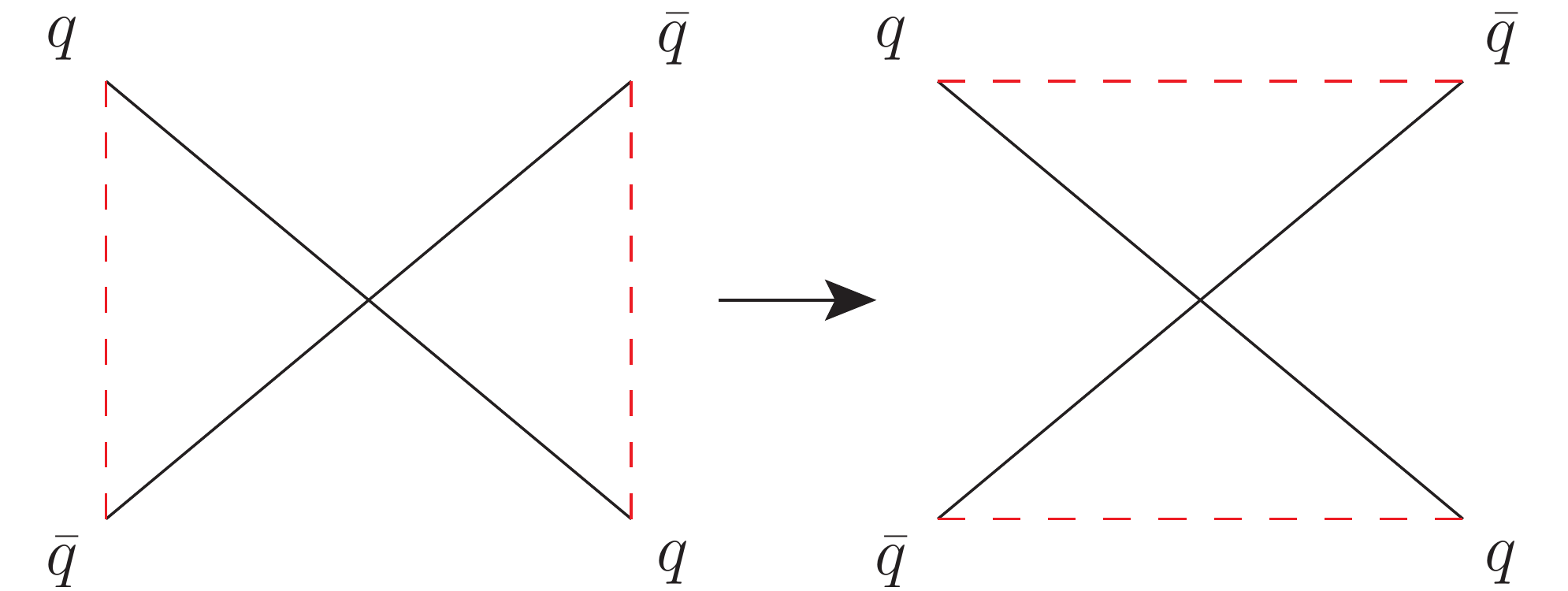} \\
  \vspace*{0.1cm}
  (a) \hspace*{7cm} (b)
  \caption{\label{fig:gluonMove}Example of the gluon-move (a) and the
    gluon-flip (b) reconnections in the gluon-move model. The dashed lines
    represent the colour configuration of the partons.}  
\end{figure*}

The gluon-move (GM) model was introduced to probe uncertainties in the 
top mass measurement, while still providing an overall good description 
of data. It is a very simple framework, in which the reduction of the 
$\lambda$ measure is at center, whereas neither colour algebra nor 
space--time geometry are considered at all.   
It contains two different types of CR, the gluon move one that gives 
the model its name, and a flip mechanism. In the former, the change in 
$\lambda$ measure is calculated if any of the gluons is moved from its
current location between two colour-connected partners to instead be
located on the string piece of any other colour-connected pair,
fig. \ref{fig:gluonMove}a. The move that lowers the total $\lambda$ measure 
the most is carried out, repeatedly until the minimum $\lambda$ measure 
is reached. The move step is quite restrictive, in that a string stretched 
between a $\q$ and a $\qbar$ endpoint will remain so; it is only the gluons 
in between that may change. Therefore an additional flip step is carried 
out after no more moves are possible. The flip mechanism flips the 
colour lines between two strings when this can reduce $\lambda$, 
fig. \ref{fig:gluonMove}b, thereby mixing up also the string endpoints 
with each other. (This is similar in character to what in another context 
is called colour swing~\cite{Bierlich:2014xba}.) 
A string is only allowed to do a single flip, to avoid 
the formation of gluon loops. The strength of the CR can be controlled by 
excluding a fraction of the gluons in the above scheme, or by requiring
the $\lambda$ reduction in a potential move/flip to be above some 
minimal value. The parameters used in this study were tuned to describe 
the LHC minimum bias data (although not quite as well as the default model). 
To allow more control, three alternative versions are considered in this
article: only including the move mechanism, GM-I, only the flip mechanism,
GM-II, and the combination of both methods, GM-III. 

The $SU(3)$-based model, CS, is similar to the GM model, in that it 
also minimizes the $\lambda$ measure by doing flips between strings. But 
it  differs in two major aspects. Firstly, it relies on the $SU(3)$ colour 
rules from QCD, together with a space--time causality requirement, to 
determine whether two strings are allowed to reconnect or not. Secondly,
it introduces a junction type of reconnection that is unique to this model.
The use of $SU(3)$ colour rules is a choice of philosophy, as already 
discussed. It limits which string pieces may flip with each other by
requiring matching colour labels, i.e.\ that the colour flow is ambiguous
already by the colour assignments of the partons. It is possible to change
the QCD-based default value, however, in the extreme case such that all
string pieces may flip with each other. For a flip between any two string 
pieces it is further required that they are in causal contact with 
each other, i.e.\ that each has had time to form before the other
has had time to hadronize. The detailed formulation of this requirement  
is ambiguous, however, so a few options are available, with a tuneable 
parameter. The appearance of junction structures offers a clear extension 
relative to the other models. An (anti)junction is a point where strings 
stretched from three (anti)coloured quarks meet. In $\e^+\e^-$ events they
must be created in pairs, one junction and one antijunction. When events
hadronize, one (anti)baryon is created around each (anti)junction, 
thereby introducing a new mechanism for baryon production. It is more 
important for high-energy hadronic collisions than it is for the 
studies in this article, however. A possibility not considered is that of
colour ropes \cite{Biro:1984cf,Bialas:1984ye,Bierlich:2014xba}, where 
several parallel strings combine into one of a higher colour representation.
If existing at all, ropes are more likely to play a non-negligible role 
in hadron or heavy-ion colliders, where the beam axis offers a natural 
alignment of many strings.

\section{W mass measurements \label{sec:mass}}

One of the key tasks of LEP2 was to determine the $\W$ mass, on its own
right and as a test of the Standard Model consistency. Measurements were 
done both in the fully hadronic and in the semileptonic
channels~\cite{Schael:2006mz,Abdallah:2008ad}. Both of them provide 
similar statistical errors, but the fully hadronic channel has a
larger systematic uncertainty, due to the CR contribution. The uncertainty 
estimate depends on the analysis method as well as on the choice of 
CR models considered (and on their parameters), but was found to be of 
the same magnitude as the statistical error. The large expected 
decrease in the statistical error at future $\e^+\e^-$ colliders
would make the fully hadronic channel irrelevant for $\W$ mass measurements,
unless the CR uncertainty could be constrained by other means. This was 
already considered at LEP2~\cite{Schael:2006mz}, where $\W$ mass measurements 
for different jet cuts were used to constrain the SK-I strength parameter.

In this section we want to turn the table, and study how a precision 
measurement of the $\W$ mass difference between the fully hadronic and the
semileptonic channels would constrain CR models and parameter values.
For this relative comparison a full optimization of both cuts and analysis 
methods is not required. Instead we will follow the method outlined in 
\cite{Sjostrand:1993hi} to provide a simple estimate of CR effects.

To this end, one million $\e^+\e^- \to \W^+\W^- \to \q_1 \qbar_2 \q_3 \qbar_4$
events were simulated for each CR model. The events are required to have 
exactly four jets using the Durham jet algorithm~\cite{Catani:1991hj}, 
with a $k_\perp$ cut of 8~GeV. In addition the jets are also required to 
have an energy  of at least 20~GeV each and be separated by an angle of 
0.5~radians. The four jets can be combined into two $\W$ bosons in three
different ways. A few options for picking the ``right'' combination are 
considered:
\begin{enumerate}
  \item With the access to MC truth information, one can try to match each
    jet with a outgoing parton of the $\W$ decays. This is done by picking the
    match that minimizes the product of the invariant masses between each jet
    and its associated parton.
  \item One can use that the $\W$ mass is known to be close to 80~GeV,
    and so minimize $|\overline{m}_{\mathrm{W}} - 80|$ to find the desired 
    match, where $\overline{m}_{\mathrm{W}}$ is the average reconstructed
    $\W$ mass.
  \item Instead of requiring the average to be close to the known $\W$ mass, 
    both masses individually could be optimized to be close to 80~GeV,
    i.e. minimize $|\overline{m}^{(1)}_{\mathrm{W}} - 80| + 
    |\overline{m}^{(2)}_{\mathrm{W}} - 80|$.
  \item At threshold the jets from the same $\W$ are almost back-to-back. 
    A match can therefore be found by maximizing the sum of opening angles.
\end{enumerate}
To a large extent these methods pick the same combinations, and thus they 
give similar results. Most of the problems arise in events with hard QCD 
radiation, where none of the methods are expected to work well. 

The $\W$ mass is calculated as the average of the two chosen $\W$ 
combinations. Since the target of this study is CR effects, the 
Breit-Wigner broadening of the mass spectrum is removed by subtracting 
the average of the produced $\W$ bosons. The results for all the methods 
are listed in table \ref{tab:massShift1}. The results for SK-I and
SK-II differ slightly from the result in the original
paper~\cite{Sjostrand:1993hi}, which is due to the $p_\perp$-ordered shower
in the newer versions of \textsc{Pythia} not being identical with the older
mass-ordered ones of the time. 

The GM model shows an interesting behaviour; the move mechanism lowers 
the $\W$ mass, while the flip mechanism increases it, and the two effects 
accidentally cancel each other in the combined result. This may be understood
as follows. If a gluon from $W_1$ is radiated at a large angle, such that 
it will move closer to the decay products from $W_2$, the move mechanism 
will connect the gluon to $W_2$, fig. \ref{fig:gluonMove}a. This will 
increase the mass of $W_2$ and decrease the mass $W_1$, but the decrease 
is larger than the increase, leading to the observed lower average mass. 
The flip mechanism instead will connect jets between the two Ws, and thereby 
increase hadronization production of particles outside the $\W$ ``cones''. 
This leads to larger opening angles, and thereby larger $\W$ masses. 
These two explanation will be revisited when studying the dedicated CR 
measurements. The complete cancellation is accidental, however, which 
becomes clear when the energy is varied. The SK-I and SK-II models also
show opposite-sign effects, thereby further stressing the message that 
the mass-shift direction of CR effects cannot be taken for granted.
Finally, the CS model shows no significant shifts, which will be
a general trend throughout all the analyses. The limitation from the colour
rules and the requirement of a lower $\lambda$ make effects very small at
$e^+e^-$ colliders. By removing the colour constraints (CS max), the model
starts to show an effect. This extreme case is already excluded at
hadron colliders, however.

\begin{table*}[t]
  \centering
  \begin{tabular}{|c|c|c|c|c|c|c|c|c|c|}
    \hline
    \multirow{2}{*}{Method} & 
    \multirow{2}{*}{$\langle \Delta \overline{m}_{\mathrm{W}} \rangle$ (MeV)} & 
    \multicolumn{8}{|c|}{$\langle \delta \overline{m}_{\mathrm{W}} \rangle$ 
    (MeV)}\\
    \cline{3-10}
    & & I & II & II$'$ & GM-I & GM-II & GM-III & CS & CS max \\
    \hline
    1 & -136 & +18  & -14  & -6  & -41 & +49  & +2 & +7  & +136  \\ 
    2 & -73  & +13  & -13  & -7  & -28 & +34  & -1 & +3  & +73  \\ 
    3 & -131 & +14  & -18  & -9  & -37 & +40  & -5 & +6  & +131  \\ 
    4 & +131 & +10  & -18  & -9  & -27 & +31  & -3 & +3  & -131  \\
    \hline
  \end{tabular}
  \caption{\label{tab:massShift1} Systematic mass shifts for the $\W$ mass 
    at 170~GeV. The $\langle \Delta \overline{m}_{\mathrm{W}} \rangle$ value
    is the average reconstructed minus produced $\W$ mass for the no-CR
    baseline. The $\langle \delta \overline{m}_{\mathrm{W}} \rangle$ is the
    additional shift for each CR model relative to this baseline. 
    The statistical uncertainty on the latter quantity is 4~MeV.}
\end{table*}

A new collider should have the capacity to increase the energy beyond the 
$\W^+\W^-$ threshold. And as was already observed for the SK-I
model~\cite{Sjostrand:1993hi}, the CR effects depend on the CM 
energy. There are two competing effects: firstly, the effect of a 
single reconnection becomes larger with increased energy, and secondly, 
the probability to have two overlapping strings decreases with energy. 
The CR mass shifts for different CM energies can be studied in
\tabRef{tab:massShift2}. Method 4 is here not included, since the 
maximum-angle method is only reliable close to the threshold. The differences
between the methods become smaller at higher energies, since the boost makes
it easier to find the right combinations. The actual shifts increase at the 
intermediate energy, but drop when the energy is increased 
further. The only model that does not show this trend is the CS model, for
which almost no effect is seen at any energy. The large shifts at the two
higher energies for the other models provide a compelling argument to 
repeat the measurements at these energies. It should be recalled, however, 
that less statistics is expected at the higher energies.

\begin{table*}[t]
  \centering
  \begin{tabular}{|c|c|c|c|c|c|c|p{0.7cm}|}
    \hline
    \multirow{2}{*}{Method}  & \multicolumn{7}{|c|}{$\langle \delta 
    \overline{m}_{\mathrm{W}} \rangle$ (MeV) ($E_{\mathrm{cm}}=$ 240 GeV)} \\
   \cline{2-8}
   &  I & II & II$'$ & GM-I & GM-II & GM-III & CS \\
    \hline
    1 & +95  & +29  & +25  & -74  & +400 & +104 & +9  \\ 
    2 & +87  & +26  & +24  & -68  & +369  & +93  & +8  \\ 
    3 & +95  & +30  & +26  & -72  & +402 & +105 & +10  \\ 
    \hline
  \end{tabular} \\
  
 \begin{tabular}{|c|c|c|c|c|c|c|p{0.7cm}|}
    \hline
    \multirow{2}{*}{Method}  &\multicolumn{7}{|c|}{$\langle \delta 
    \overline{m}_{\mathrm{W}} \rangle$ (MeV)  ($E_{\mathrm{cm}}=$ 350 GeV)} \\
    \cline{2-8}
   &  I & II & II$'$ & GM-I & GM-II & GM-III & CS \\
    \hline

    1  & +72  & +18  & +16  & -50 & +369 & +60 & +4  \\ 
    2  & +70  & +18  & +15  & -50 & +369 & +60 & +4  \\ 
    3  & +71  & +18  & +16  & -50 & +369 & +60 & +3  \\ 
    \hline
  \end{tabular}
  \caption{\label{tab:massShift2} Systematic $\W$ mass shifts 
    at center-of-mass energies of 240 and 350 GeV, respectively.
    The $\langle \delta \overline{m}_{\mathrm{W}} \rangle$ is the
    mass shift in the CR models relative to the no-CR result. 
    The statistical uncertainty is 5~MeV.}
\end{table*}

\section{Four-jet angular distributions \label{sec:angular}}

The direct searches for CR in $\W^+\W^-$ events at LEP ruled out  
extreme parameter values for SK-I and potentially could also rule out 
some of the new CR models. Especially the GM-I and GM-II models have 
that potential, since they were already observed to have a larger effect 
on the $\W$ mass measurement than the other models.

The analysis relies on the particle multiplicities in the angular regions 
between two jets from the same $\W$ decay and from different $\W$ decays,
respectively, to provide a ratio that is sensitive to CR. The idea is that 
a reconnection will form a string between jets from different $\W$ decays, 
thereby increasing the multiplicity between those jets. In general, we will 
therefore expect the same-to-different ratio to become lower when CR is 
switched on. Several LEP
experiments~\cite{Achard:2003pe,Abdallah:2006uq,Abbiendi:2005es} performed
this measurement. The results presented in the studies are after detector 
simulation, however, and as such are not directly comparable with the 
results obtained in this study. Instead we will rely on the ratio between 
the CR and the no-CR results ($r$, see later for exact definition), since 
detector effects are reduced for this observable. A preliminary combination 
of the different experiments gave $r = 0.969 \pm 0.011 (\mathrm{stat.}) \pm 0.009
(\mathrm{syst. corr.}) \pm 0.006 (\mathrm{syst. uncorr.})$~\cite{Alcaraz:2006mx}
corresponding to a 2.2 standard deviation disagreement with the no-CR scenario.
A later combined study~\cite{Schael:2013ita} has increased this to disfavor 
the no-CR model at a 2.8 standard deviation level, by combining with the 
mass shift results and performing a $\Delta \chi^2$ fit. No separate $r$ 
results were shown, however, and therefore we will have to rely on the 
preliminary combination.

\begin{figure}
  \centering
  \includegraphics*[scale=0.7]{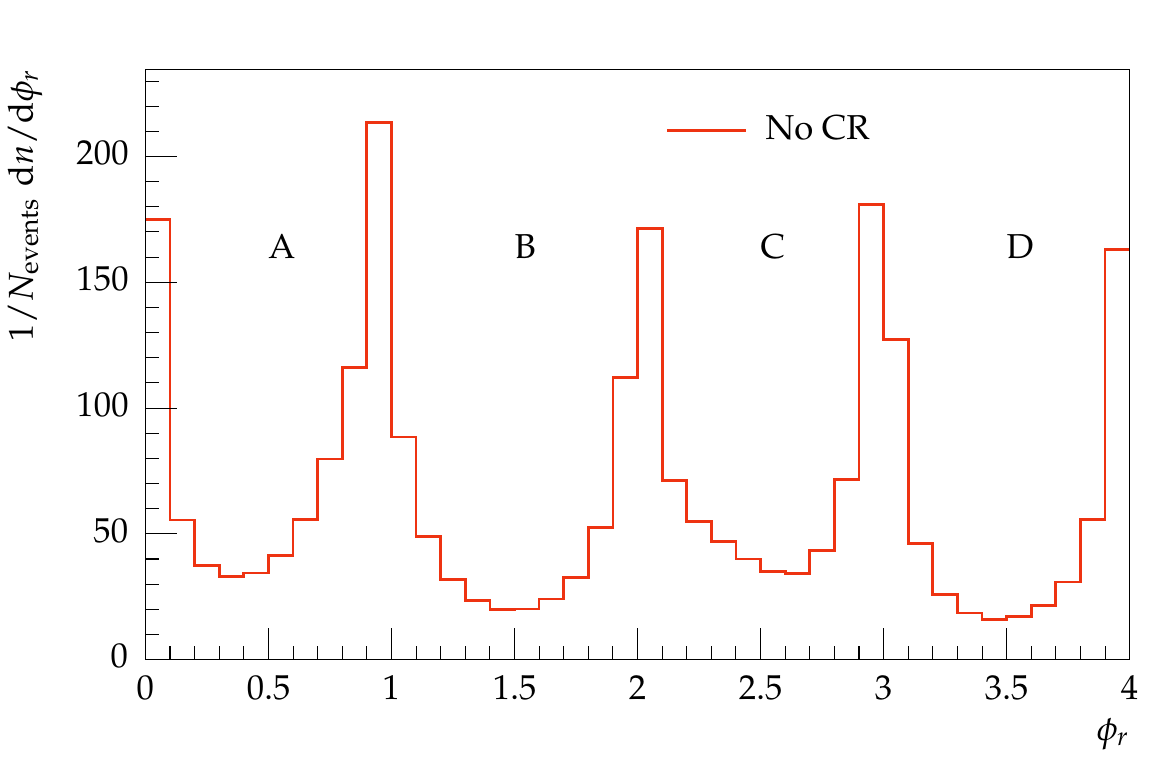}
  \caption{\label{fig:angle} The $\varphi_r$ distribution at a center-of-mass
    energy of 183 GeV.}
\end{figure}

The event selection and analysis procedure varied slightly between the
different LEP experiments. Two of the experiments relied purely on the angles
to pair the jets~\cite{Achard:2003pe,Abdallah:2006uq}, while one experiment 
also used the invariant masses~\cite{Abbiendi:2005es}. We decided to mimic 
the analysis from the L3 collaboration~\cite{Achard:2003pe}. A short recap 
of the event selection and analysis is presented, but for more details we 
refer to the experimental studies. 

The event selection requires each event to have exactly four jets with the 
Durham jet algorithm, with $y_\mathrm{cut} = 0.005$. The two smallest 
of the six interjet angles are required to be below $100^{\circ}$ and be
non-adjacent. These are assumed to be the two regions between the different 
$\W$ decays, and are normally referred to as regions B and D. In addition
two more angles are required to be between $100^{\circ}$ and $140^{\circ}$ and 
be non-adjacent. These are assumed to be the regions inside the $\W$ decays,
and are normally referred to as region A and C. If several combinations 
are allowed, the one with the largest total opening angle is chosen. 
For each region the particles are projected onto the plane spanned by the 
two jets, and all particles are assigned a rescaled angle 
$\varphi_r = \varphi/\varphi_{\mathrm{jj}}$, where $\varphi$ is the angle 
from the particle to one of the jets and $\varphi_{\mathrm{jj}}$ is the angle 
between the two jets. This distribution is shown in \figRef{fig:angle}, 
where the different regions are separated by adding an integer to each. 
The final observable is defined as 
\begin{equation}
  R_N =
  \frac{\int_{0.2}^{0.8}\frac{dn}{d\varphi_r}(A+C)d\varphi_r}
  {\int_{0.2}^{0.8}\frac{dn}{d\varphi_r}(B+D)d\varphi_r}
  \, .
\label{eq:RN}
\end{equation}
The regions closest to the jets are excluded since they are mainly sensitive 
to the internal jet evolution. Finally the ratio between the different 
CR models and the no-CR baseline is defined as 
$r = R_N^{\mathrm{CR}} / R_N^{\mathrm{no CR}}$. Thus a deviation from unity would 
disfavour the no-CR scenario. The results for the
various CR schemes are shown in tab.~\ref{tab:fourjet1}. As expected, 
all CR models, except for GM-I, predicts an $r$ below unity. The GM-I model 
only allows the gluon move reconnections, and therefore it does not reconnect 
the quarks at the string endpoints. Instead, it can take gluons emitted 
at large angles and move them to the other $\W$ string,
thereby actually lowering the amount of radiation in region B and D,
fig. \ref{fig:gluonMove}a. This is the same explanation as for the lower
$\W$ mass observed in section \secRef{sec:mass}. The GM-II model only does 
flips, which is exactly what this observable is optimized to measure. 
This is in fair agreement with observations, since this model shows relative 
large deviations from unity. The SK-I model with default strength is quite 
well in agreement with the actual measurement. For comparison the maximal 
SK-I model, where a reconnection is always done, is also included. 
It gives too large shifts and so can be excluded. The SK-II models and 
the CS model do not produce any large shifts. The maximal CS model,
where the $SU(3)$ rules are ignored and CR is only limited by the $\lambda$ 
measure, shows a larger effect and it can potentially be ruled out by
experiments. It is, however, still relatively small compared to the other
maximal CR models.

In this study we consider several intervals, and not only the 0.2--0.8 
considered in the original study. A clear trend shows that the smaller 
the interval, the more sensitive the observable becomes, i.e. varies more
from unity. This is not surprising since the region closest to the jets 
are dominated by their perturbative behaviour. It should be noted that the 
statistics becomes worse for smaller intervals, but with the larger 
expected statistics at a new collider, a smaller interval than at LEP2
would most likely be preferable.

\begin{table*}[t]
  \centering
  \begin{tabular}{|c|c|c|c|c|c|c|c|c|c|c|}
    \hline
   inter- & \multirow{2}{*}{$R_N^{\mathrm{no CR}}$} & \multicolumn{9}{|c|}{$r$}\\
    \cline{3-11}
    val& & I & II & II$'$ & GM-I & GM-II & GM-III & CS & I max & CS max \\
    \hline
    0.1--0.9 & 1.1031 &     0.9889 &     0.9971 &     0.9969 &     1.0132 &
    0.9629  & 0.9876 & 0.9960 &     0.9614 &     0.9712 \\
    0.2--0.8 & 1.1482 &     0.9802 &     0.9916 &     0.9931 &     1.0293 &
    0.9440  & 0.9918 &    0.9910 &     0.9360 &     0.9781 \\ 
    0.3--0.7 & 1.1402 &     0.9747 &     0.9887 &     0.9889 &     1.0404 &
    0.9301  & 0.9931 &     0.9911 &     0.9196 &     0.9831 \\ 
    0.4--0.6 & 1.0883 &     0.9702 &     0.9823 &     0.9880 &     1.0460 &
    0.9181  & 0.9882 &     0.9920 &     0.9068 &     0.9810 \\ 
    \hline
  \end{tabular}
  \caption{\label{tab:fourjet1} Results for $R_N$ and $r$ for different
    intervals in $\varphi_r$ at a center-of-mass energy of 183 GeV. Two maximal
    CR models are included for SK-I and for the QCD based method,
    respectively. The statistical uncertainty on $r$ is around 0.0025. }
\end{table*}

To check if the new models are already excluded by the LEP measurements, the
number of standard deviations from the measured result is calculated,
\tabRef{tab:fourjet2}. The experimental uncertainties are assumed Gaussian
and added in quadrature. The only model excluded at the three $\sigma$ level
is the GM-I model, which is the only model predicting a larger than unity
$r$. The uncertainty is still too large to invalidate the other models, and a
new collider with higher precision is needed to constrain these.

\begin{table*}[t]
  \centering
  \begin{tabular}{|c|c|c|c|c|c|c|c|c|c|c|c|}
    \hline
    & no CR & I & II & II$'$ & GM-I & GM-II & GM-III & CS & I max & CS max \\
   \hline
    $n_\sigma$ & 2.0 & 0.7 & 1.5 & 1.6 & 3.9 & 1.6 & 1.5 & 1.4 & 2.1 &
    0.6 \\
    \hline
  \end{tabular}
   \caption{\label{tab:fourjet2} Deviations from the measured result shown in
     number of standard deviations 
     ($n_\sigma=(r_\mathrm{exp}-r_\mathrm{th})/(\delta r)_\mathrm{exp}$).}
\end{table*}

The $\W$ mass measurement was seen to be more sensitive to CR at higher 
energies, and hence a similar effect is expected here. The method described 
above cannot directly be applied at higher energies, however, since the 
increased boost of the $\W$ bosons changes the angular distributions
between the jets. Instead we apply a method similar to method 3 in the 
$\W$ mass section to define the two angles within the $\W$ decays. The 
two other angles are defined to minimize the total sum of their angles. The 
results for the different energies are shown in table \ref{tab:fourjet3}. 
The new method performs slightly worse at 183~GeV, i.e.\ the ratios lie 
closer to unity. This is especially evident when considering the 
maximal CR models. At higher energies, however, the deviation from unity
becomes larger for some of the more extreme models, indicating a better
sensitivity, but this observable shows no sensitivity for the CS
model. The moderate models do not show any significant variation with 
energy, and as such it is difficult to tell whether the potential limits 
on CR can be stronger at higher energies. In general we expect a falling 
fraction of events with CR for higher energies, but more spectacular effects
for the events where CR occurs, so in the future we will need to search 
for more selective tests.

\begin{table*}[t]
  \centering
  \begin{tabular}{|c|c|c|c|c|c|c|c|c|c|c|}
    \hline
    \multirow{2}{*}{$\sqrt{s}$ [GeV]} & \multirow{2}{*}{$R_N^{\mathrm{no CR}}$} 
    & \multicolumn{9}{|c|}{$r$}\\
    \cline{3-11}
    & & I & II & II$'$ & GM-I & GM-II & GM-III & CS & I max & CS max \\
    \hline
    183     & 1.9003 &     0.9900 &     0.9915 &     0.9924 &     1.0142 &
    1.0247 &     0.9768 &     0.9902 &     0.9667 &     1.0147 \\ 
    240     & 1.1764 &     0.9820 &     0.9935 &     0.9933 &     0.9857 &
    1.0130 &     0.9362 &     0.9993 &     0.9030 &     1.0006 \\ 
    350     & 1.4459 &     0.9829 &     0.9948 &     0.9939 &     0.9758 &
    1.0022 &     0.9228 &     1.0028 &     0.8502 &     0.9946 \\ 
    \hline
  \end{tabular}
  \caption{\label{tab:fourjet3} Results for $R_N$ and $r$ for different
     center-of-mass energies for a fixed interval (0.2--0.8). The statistical
     uncertainty on $r$ is around 0.0015. } 
\end{table*}

As a slightly simpler observable, to test CR, it is possible to study the
overall multiplicity. In most models CR minimizes the $\lambda$ measure 
and therefore also
lowers the total multiplicity. This is normally compensated by a retuning of
the hadronization parameters or the perturbative regime. But by comparing the
multiplicity in fully hadronic and semileptonic $\W^+\W^-$ events, it is
possible to directly probe CR. If no CR is switched on, the ratio
$N_{\mathrm{ch}}^{\W^+\W^- \to \q_1 \qbar_2 \q_3 \qbar_4} / 
(N_{\mathrm{ch}}^{\W^+\W^-  \to \q_1 \qbar_2 \ell \nu_{\ell}} -1) $ is expected 
to be exactly equal to 2 (with $\ell = \e$ or $\mu$, but excluding $\tau$).
A simple study at a center-of-mass of 170 GeV shows that indeed it is 
interesting to use this observable. Both the individual GM models show
an effect, 1.96 and 1.97 for GM-I and GM-II, respectively. Contrary to the
earlier observables, the two effects add coherently and the combined result is
1.93. With 1.97 the CS model also shows more sensitivity in this observable as
compared to the more complicated four-angle measurement. Similar results are
also obtained for the SK models, so this would be an intriguing
measurement for a future $\e^+\e^-$ collider.

\section{Higgs parity  measurements \label{sec:higgs}}

As discussed in the introduction, hadronic $\W^+\W^-$ and $\Z^0\Z^0$ decays 
of the 125~GeV Higgs offers a novel system for CR effects. Like in the 
$\W^+\W^-$ studies above we should not expect big effects, so it is unlikely
to be discernible in the busy LHC environment. In a process like 
$\e^+\e^- \to \gamma^*/\Z^{0*} \to \H^0\Z^0 \to \H^0\ell^+\ell^-$, 
or $\mu^+\mu^- \to \H^0$ for that matter, detailed studies should become 
possible, however, assuming sufficient luminosity. As before, reconstructed 
masses and angles may become affected. Rather than simply repeating 
discussions along the lines of the previous two sections, we choose to 
illustrate possible effects for another set of observables, related to 
setting limits for  $CP$ violation in Higgs decays. We are aware that 
such tests can be performed in purely leptonic decays, say 
$\H \to \Z^0\Z^0 \to \mu^+ \mu^- \e^+ \e^-$, although with a much lower
branching ratio. It can also be probed by the decay angles of the 
$\Z^0$ produced in the association with the $\H^0$ \cite{Skjold:1995jp}. 
The purpose of this brief study is not to compare the relative merits 
of $CP$-violation tests in these different channels, but to stay with
$\H \to \W^+\W^- \to \q_1 \qbar_2 \q_3 \qbar_4$ and check what CR could
mean there. To this end we will use a simplistic $\chi^2$ test on what
could be the most sensitive variable. 

To simulate a mixed $CP$-even and $CP$-odd Higgs boson, we will use the 
Higgs doublet model already implemented in \textsc{Pythia}, with the option
to allow $CP$-violation based on the expressions in \cite{Skjold:1993jd}. 
We will assume that the 125~GeV Higgs is almost completely $CP$-even, 
with a small admixture of $CP$-odd. Allowing for an interference term 
between the two, the Higgs cross section can be written as
\begin{equation}
  \sigma \propto k_{\mathrm{even}}^2 A + k_{\mathrm{odd}}^2 B +
  k_{\mathrm{even}}k_{\mathrm{odd}} C \, ,
\end{equation}
where $A,B,C$ depends on the kinematics of the event and the $k$ determine 
the contributions to the different types. Since $A, B$ and $C$ are not of 
the same order of magnitude, a characterization in terms of a mixing angle
is not convenient. Instead we use a definition based on the fraction, 
later referred to as parity fraction, of the events coming from either of 
the odd and the interference parts of the cross section:
\begin{equation}
  f = \frac{|k_{\mathrm{odd}}^2 B| + |k_{\mathrm{even}}k_{\mathrm{odd}} C|}
    {|k_{\mathrm{even}}^2 A| + |k_{\mathrm{odd}}^2 B| +
    |k_{\mathrm{even}}k_{\mathrm{odd}} C| } \, .
\end{equation}
For an almost $CP$-even Higgs, $f=0$, this quantity provides a reasonable 
estimate of the amount of $CP$-violating interference introduced for the 
Higgs boson. 

The parity of the Higgs can be measured by studying the angles between the
fermions from the boson decays. In the standard analyses of the spin/parity 
of the Higgs boson (see e.g.~\cite{Gao:2010qx,Bolognesi:2012mm}), five such
angles are defined, out of which three are 
sensitive to the parity of the Higgs. These three angles are: $\theta_1$, 
the polar angle of a fermion in the rest frame of its $\W$ mother, with
respect to the direction of motion of the $\W$ in the $\H$ rest frame, 
$\theta_2$, similarly but for the other $\W$, and $\Phi$, the angle between 
the two planes spanned by the decay products of the respective $\W$ bosons. 
The rest of this section will therefore be a study on the effect of CR on 
these three angles. 

\begin{figure*}[th]
\centering
\includegraphics*[scale=0.6]{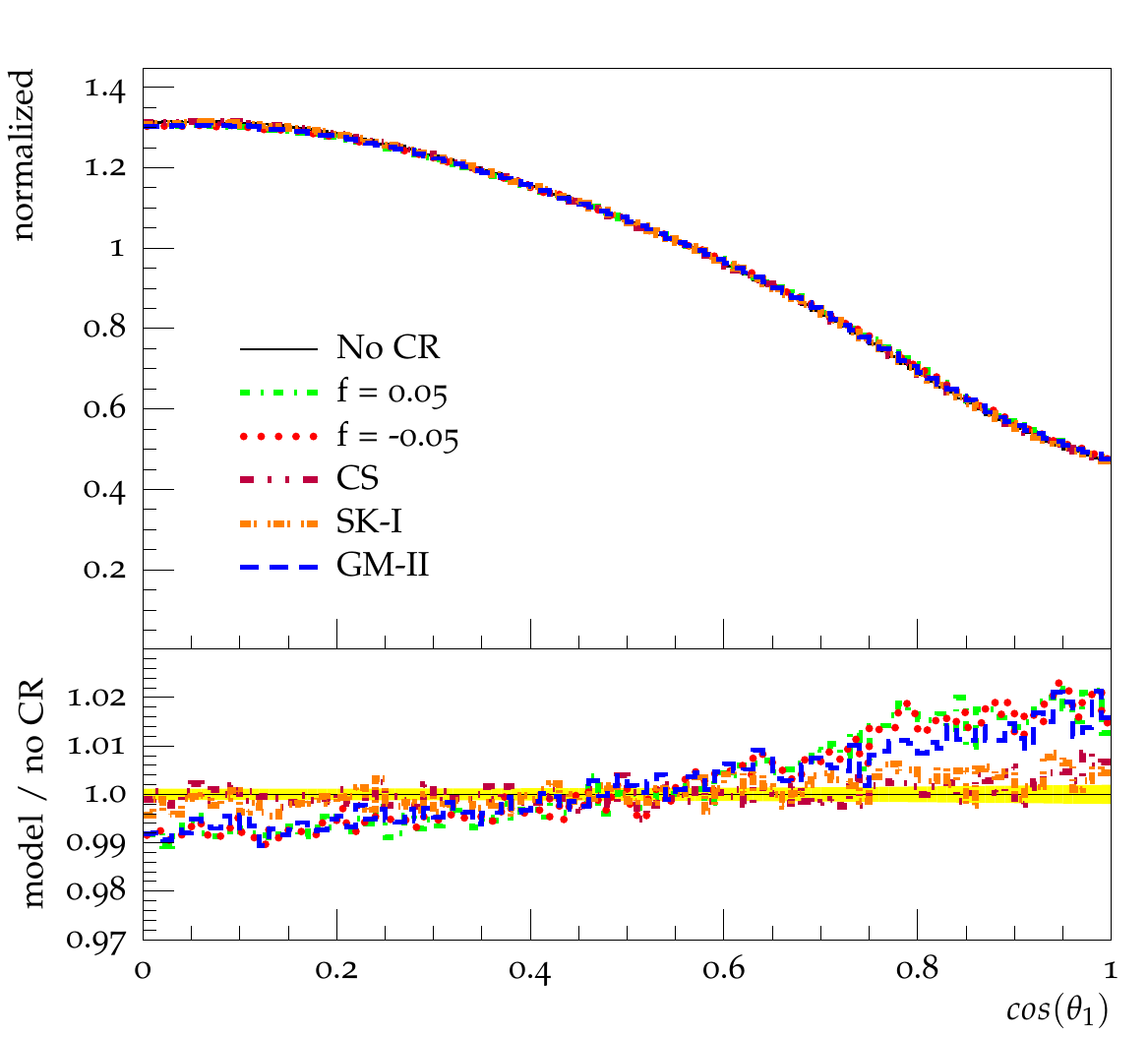}
\includegraphics*[scale=0.6]{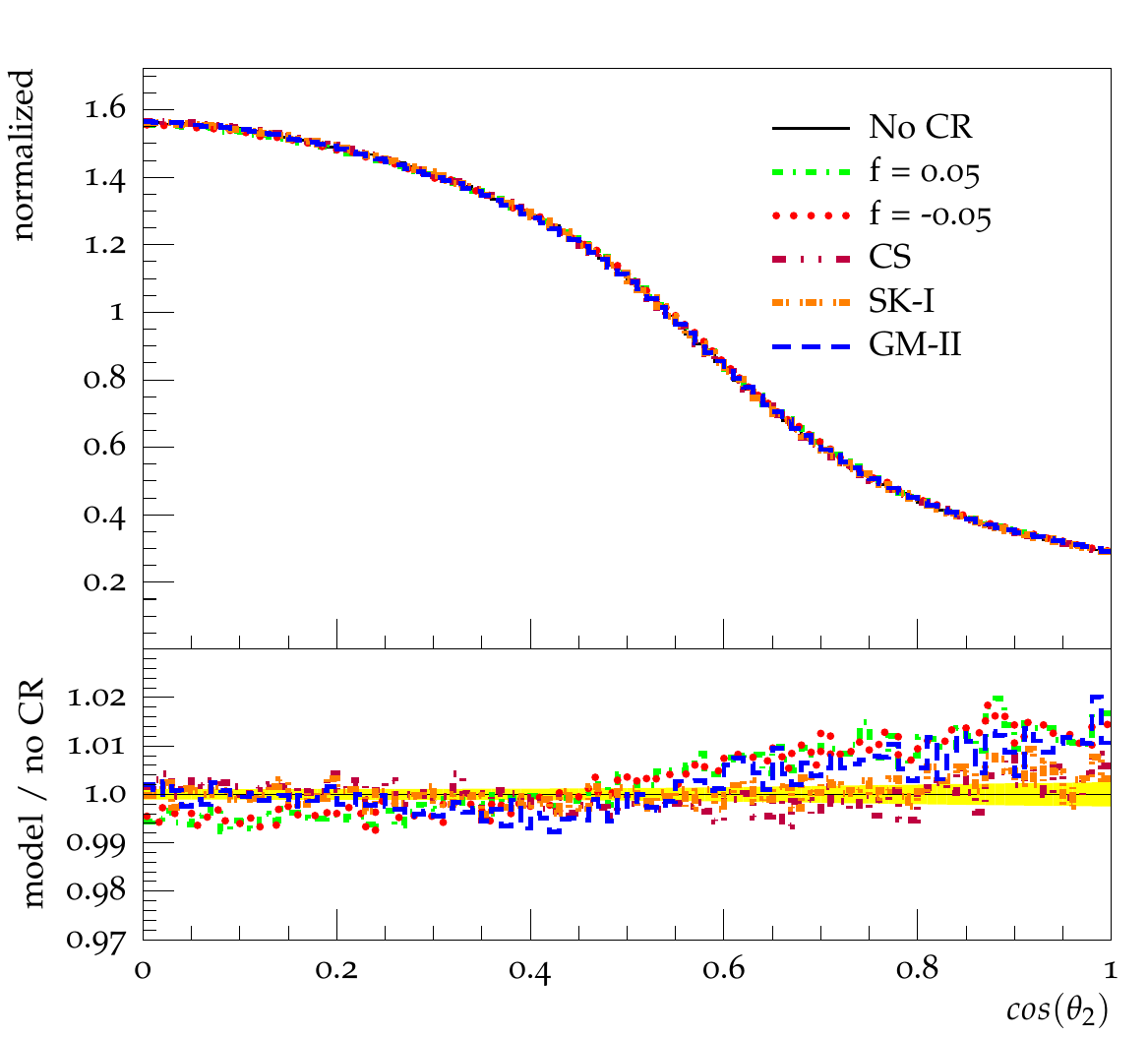} \\
(a) \hspace*{7cm} (b) \\
\includegraphics*[scale=0.6]{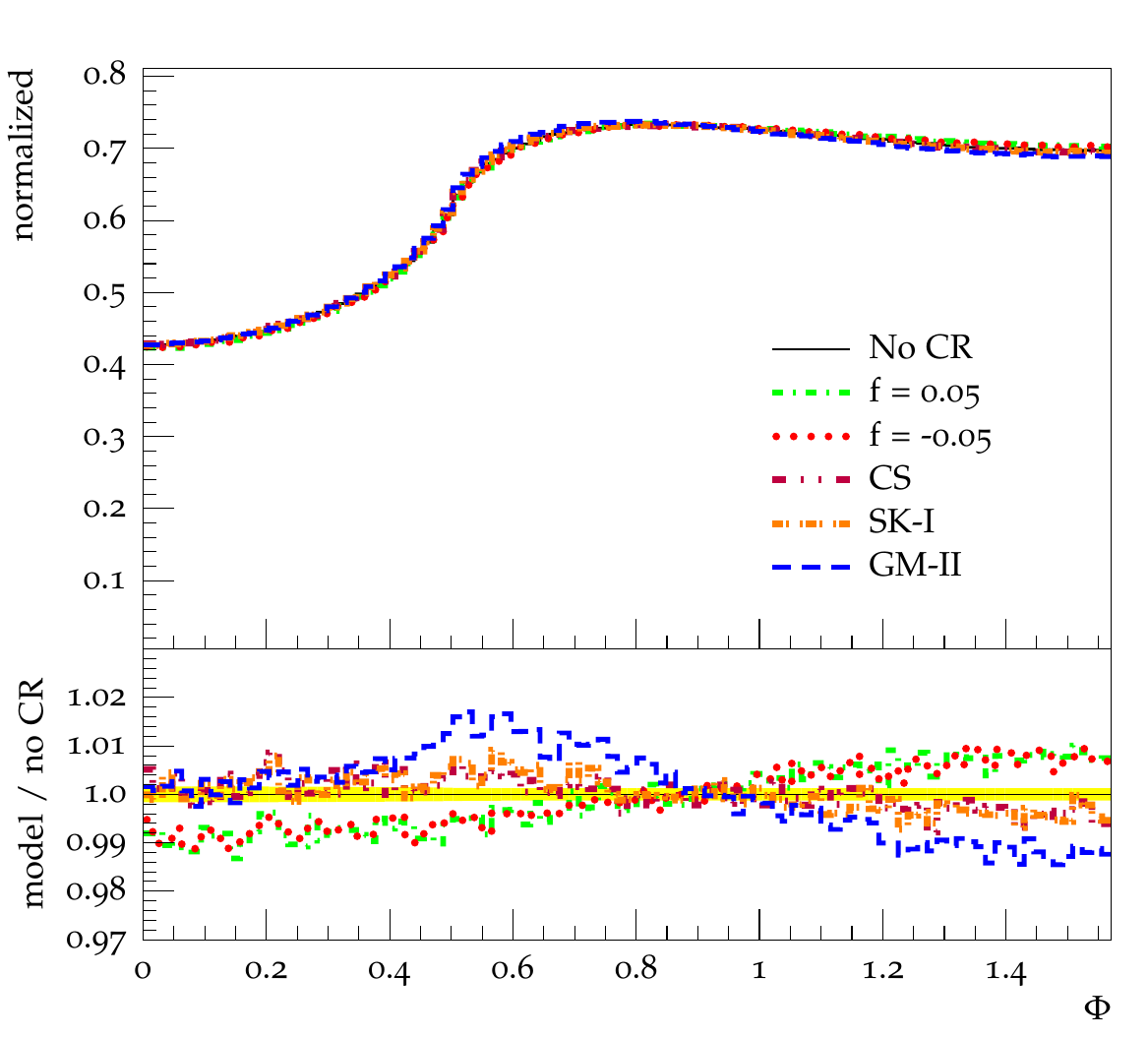} \\
(c)
\caption{The three angles sensitive to the parity of the Higgs boson. Three
  different parity scenarios are shown together with a small selection of
  different CR models.
\label{fig:parAngles}}
\end{figure*}

To only have to consider the Higgs decay itself we have studied the process
$\mu^+\mu^- \to \H^0$, but this should only be viewed as a technical trick.
All models are set up to easily handle this, whereas $\e^+\e^- \to \H^0 \Z^0$ 
would require a bit more bookkeeping for the SK models. Otherwise the models
remain unchanged relative to previous studies. The fact that at least one of 
the $\W$'s have to be strongly off-shell implies that its lifetime is 
considerably reduced, and this is taken into account in the SK models.   
To estimate the effect of CR on the angles, 100 million 
$\mu^+\mu^- \to \H^0 \to \W^+\W^- \to \q_1\qbar_2\q_3\qbar_4$ events 
are simulated for each CR model and for each parity fraction, respectively. 

The events are required to have exactly four jets using the Durham jet 
algorithm with a $k_\perp$ cut of 8~GeV, followed by an additional energy 
cut of at least 10~GeV per jet and a angular separation of 0.5. Two 
different methods to pair the jets were considered, either to maximize 
the opening angles, or to minimize $|M_{\W} - 80|$ for a single $\W$. 
The second method was found to be significantly more sensitive, and we will 
therefore restrict ourselves to this method. The distribution for the three 
angles are shown in \figRef{fig:parAngles}. Deviations between the SM Higgs 
and the different parity fractions are visible by eye for all the three 
angles. Both of the curves with nonvanishing $CP$-oddness show almost 
identical behaviours, indicating that the sign of the interference term 
is unimportant for these observables (at least for small deviations). 
Comparing the pattern of variation for the $CP$-violating models and the 
CR models, respectively, shows an interesting picture. For 
$\theta_1$ and $\theta_2$ the deviations go in the same direction, 
whereas for $\Phi$ the deviations are in opposite directions. Thus a
simultaneous study in principle would allow to disentangle the two
potential effects.

\begin{figure}[t]
  \centering
  \includegraphics*[scale=0.7]{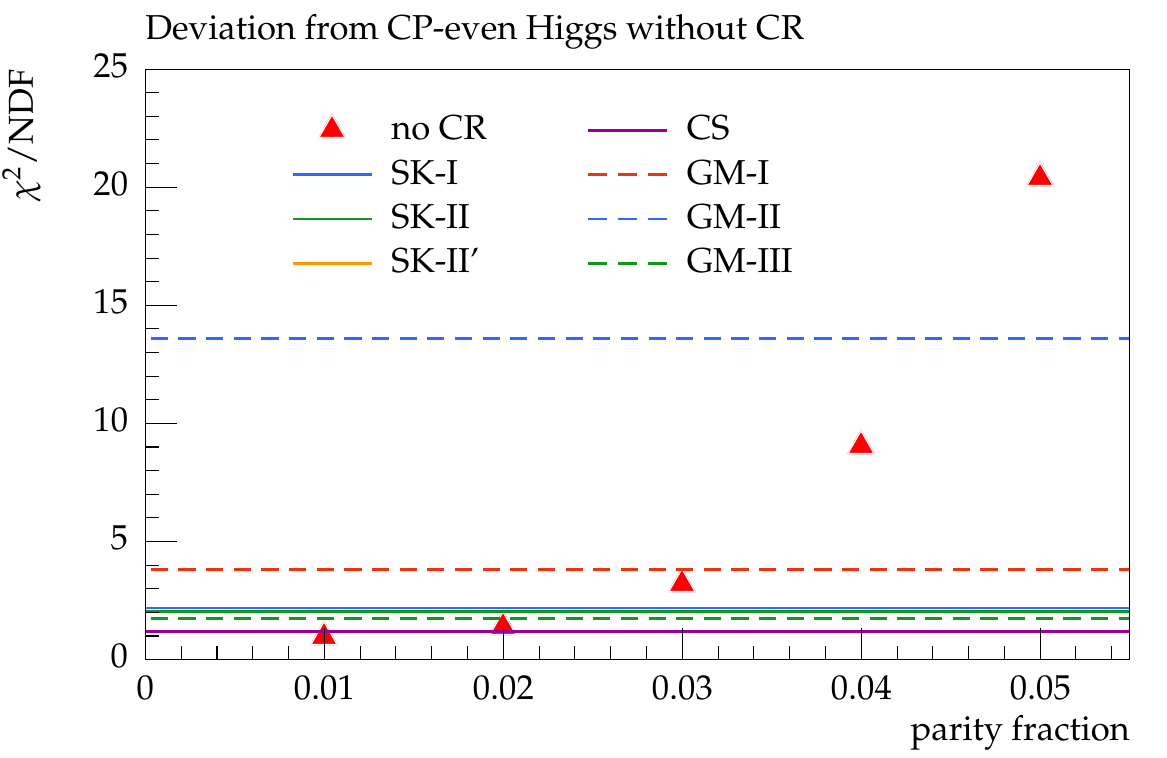}
\caption{Deviations between a $CP$-even Higgs without CR and models with either
  increased $CP$-oddness or a CR model. The deviation is quantified as the
  $\chi^2/\mathrm{NDF}$ deviation for the $\varphi_1$ angle.
\label{fig:chi2}}
\end{figure}

To quantify the deviation from the no-$CP$-odd no-CR baseline, a simple 
$\chi^2$ test is applied to the distributions. The most sensitive angle 
is $\theta_1$, and we therefore restrict our studies to this observable. 
A complete experimental analysis most likely would combine all the angles 
in a multivariate analysis. For each parity fraction the $\chi^2$ is 
calculated, fig.~\ref{fig:chi2}. As expected the $\chi^2$ increases smoothly 
with this fraction. Similarly, the $\chi^2$ is also included for the 
different CR models. The crossover point is a simple indicator for when 
CR becomes an issue for Higgs parity measurements. This point occurs 
around 2--5\%, with the higher values for somewhat more extreme CR models.
Thus any limits significantly above this estimate can safely ignore CR 
effects. It should be stressed that also limits below 2\% should be
reachable, once CR is carefully taken into account. This can involve 
(anti)correlations between the three angles, as already noted, but also
studies of particle production patterns between the jets, like the one 
in section \ref{sec:angular}. 

\section{Conclusions \label{sec:conclusions}}

In this article we have studied the effects of CR at $\e^+\e^-$ colliders,
with emphasis on fully hadronic $\W^+\W^-$ events. We find that some newer
models, implemented to study CR effects at hadron colliders, show different
behaviours for $\e^+\e^-$. The CS model gives rise to very limited variations,
whereas for the GM models one specific scenario even shows large enough
deviations to be excluded by the LEP data. 

Even if the concept of CR is quite straightforward, it allows for several 
different mechanisms to be at play. These potentially act in opposite 
directions, making interpretations difficult. This is clearly illustrated 
by the GM models,  where GM-I predicts a smaller reconstructed $\W$ mass
and GM-II a larger one. This highlights the need for studying multiple 
models using several observables, to disentangle what is going on. 
Much further work is needed, but the outcome of the current simple study 
is fairly optimistic: given enough luminosity, at a few different energies, 
$\e^+\e^-$ should offer insights into CR mechanisms that complement those 
obtainable at hadron colliders. This complementarity between the ``clean''
$\e^+\e^-$ environment and the ``dirty'' $\p\p$ one may hold the key to a 
deeper understanding of CR. 

The $\e^+\e^- \to \W^+\W^-$ channel is not the only $\e^+\e^-$ process where  
CR effects may be relevant. As an example we studied a Higgs parity
measurement in the $\H \to \W^+\W^- \to \q_1 \qbar_2 \q_3 \qbar_4$ channel. 
The variations from CR were of the same size as the introduction of 2--5\%
$CP$-oddness into the $CP$-even Higgs, depending on the choice of CR model. 
The main lesson is not the precise number for this particular observable, 
but to highlight the need to be aware of potential CR uncertainties for
any nontrivial hadronic final state.

Plans for future $\e^+\e^-$ collider usually include the possibility to 
reach the $\t \tbar$  threshold. Then hadronic final states will start out 
with three colour singlets: one $\W$ from each top decay, plus one 
encompassing the $\b$ and $\bbar$ from the two decays. Like for the   
$\W^+\W^-(\gamma^*/\Z^0)$ background this increases the possibilities for
CR effects. Some early studies are found in~\cite{Khoze:1994fu}, but 
updated and extended studies should be performed, including the new models. 
At the very least, it will be needed in order to estimate the expected 
CR uncertainty in the measurements of the top properties for possible 
future colliders. Many of the necessary tools are already in place in
\textsc{Pythia}~8, although e.g.\ the administrative machinery in the
SK models needs to be extended appropriately.

\section*{Acknowledgments}

Work supported in part by the Swedish Research Council, contract number
621-2013-4287, and in part by the MCnetITN FP7 Marie Curie Initial 
Training Network, contract PITN-GA-2012-315877. Peter Skands is acknowledged
for useful comments.

\clearpage
\bibliographystyle{utphys}
\bibliography{article}

\end{document}